\documentclass[a4paper,11pt]{article}
\usepackage{pos}

\title{Flipped rotating axion: Baryogenesis and Dark Matter}

\author*[a]{Konstantinos Dimopoulos}

\affiliation[a]{Consortium for Fundamental Physics, Physics Department, Lancaster University,\\ Lancaster LA1 4YB, United Kingdom}

\emailAdd{k.dimopoulos1@lancaster.ac.uk}

\abstract{It is shown that the co-genesis of baryon asymmetry and dark matter can be achieved through the rotation of a spectator axion-like particle, because of a flip in the vacuum manifold's orientation at the end of inflation. This can occur if the axion has a periodic non-minimal coupling to gravity (while preserving the discrete shift symmetry) in non-oscillating inflation models, where the inflaton field is characterised by a runaway potential. Our rotating axion can generate the baryon asymmetry of the Universe through spontaneous baryogenesis, while at a later epoch it can oscillate as dark matter. We show that in order to avoid fragmentation of the axion condensate during the rotation, we require the non-minimal coupling \mbox{$\xi \sim (f/m_P)^2 $}, where $f$ is the axion decay constant.}

\FullConference{Proceedings of the Corfu Summer Institute 2025 "School and Workshops on Elementary Particle Physics and Gravity" (CORFU2025)\\
27 April - 28 September, 2025\\
Corfu, Greece\\}

\begin{document}
\maketitle

\section{Introdcution}

The Universe is homogeneous and isotropic, states the Cosmological Principle. If the principle were exact,
the Universe should have been filled only with a thinned out thermal radiation at 3~K. 
Instead it is filled with structures, such as galaxies, made of matter. And not anti-matter!

When matter meets anti-matter it annihilates to radiation.
Consequently, there must by an imbalance between matter and anti-matter. This imbalance is called the
Baryon Asymmetry.
The baryon asymmetry is tiny:
\mbox{$Y_B=n_B/n_r=8.7\times 10^{-11}$}, where $n_B$ is the number density of the usual baryonic matter and $n_r$ is the number density of radiation in the Universe, which could be the product of annihilation of matter and anti-matter.
Even though \mbox{$Y_B\ll 1$}, the question remains:
Where is all the anti-matter?
It seems we need a mechanism that
accounts for the baryon asymmetry. Such a process is called Baryogenesis.

In addition, more than 80\% of matter is invisible. We call it Dark Matter (DM). It is usually assumed that DM is either
Weekly Interacting Massive Particles, or
Primordial Black Holes, or Axions. We will focus in the last possibility \cite{Co:2019wyp}.
Axions are ultralight scalar particles, which satisfy a shift-symmetry. They were
originally introduced to solve the strong CP problem of QCD \cite{Peccei:1977hh}, i.e. nothing to do with DM.
Axions are ubiquitous in fundamental theory, e.g. the
String Axiverse \cite{Arvanitaki:2009fg}.

In this work \cite{Chen:2025awt},
with collaborators 
Chao Chen, Suruj Jyoti Das and Anish Ghoshal, we introduce an axion, non-minimally coupled with gravity, rotating in field space, that can lead to Baryogenesis and become the Dark Matter at present.
In the following we use natural units, for which \mbox{$\hbar=k_B=c=1$} and \mbox{$8\pi G=m_P^{-2}$}, with \mbox{$m_P=2.43\times 10^{18}\,$GeV}
being the reduced Planck mass.  

\section{The flipped rotating axion}

The Lagrangian density of the model is
\begin{equation}
	{\cal L}=\frac12 m_P^2\gamma^2(\phi)R-
	\frac12 
	(\partial\phi)^2-V(\phi)
	\label{L}
\end{equation}
introduced in Refs.~\cite{Ferreira:2018nav,Salvio:2021lka,Ghoshal:2023jvf}. 
In the above, \mbox{$(\partial\phi)^2\equiv\partial_\mu\phi\,\partial^\mu\phi$} and 
\mbox{$\gamma^2(\phi)\equiv 1+\xi\left[1-\cos(\phi/f)\right]$},
where the unity in the right-hand side denotes the usual Einstein-Hilbert term, and
\begin{equation}
	V(\phi)=M^4\left[1-\cos(\phi/f)\right]\,,
	\label{V}
\end{equation}
where $\xi$ the non-minimal coupling of the axion 
field $\phi$ to the Ricci scalar $R$, $f$ is the axion decay constant with \mbox{$0<f\ll m_P$} and $M$ is the 
symmetry breaking scale with \mbox{$0<M\ll f$}. The above Lagrangian density shows that the theory respects the discrete shift symmetry \mbox{$\phi\rightarrow\phi+2\pi f$}.  

Eq.~\eqref{L} can be written as
\begin{equation}
	{\cal L} = \frac12m_P^2 R-\frac12(\partial\phi)^2 -\left(M^4-\frac12\xi m_P^2 R\right)[1-\cos(\phi/f)] ~.
	\label{Lphi}
\end{equation}
When \mbox{$\xi\ll 1$}, the non-minimal coupling term is much smaller than the Einstein-Hilbert term, but it can be compared with $M^4$ that can be very small.  Therefore, we are effectively in Einstein gravity with the non-minimal coupling practically only contributing to the effective potential.

In FRW spacetime \mbox{$R=3(1-3w)H^2$}, where $w$ and $H$ are the barotropic and the Hubble parameter, respectively. We consider that our axion is a spectator field in a non-oscillatory inflationary scenario (a runaway inflaton
field is assumed), where inflation (with \mbox{$w=-1$}) is followed by a period of kination (with \mbox{$w=1$}). Thus, the non-minimal contribution changes sign from inflation to kination afterwards, as with Ricci reheating \cite{Dimopoulos:2018wfg,Opferkuch:2019zbd,Bettoni:2021zhq}.

Assuming
\mbox{$M^4\ll\xi m_P^2R$}, 
during inflation, the axion effective potential is 
\begin{equation}
	V_{\rm eff}(\phi)\simeq -6\xi m_P^2 H^2[1-\cos(\phi/f)] ~,
	\label{Veffinf}
\end{equation}
while during kination we have
\begin{equation}
	V_{\rm eff}(\phi)\simeq 3\xi m_P^2 H^2[1-\cos(\phi/f)] ~,
	\label{Veffkin}
\end{equation}
where we have ignored $M$.

\begin{figure*}[t]
	\centering
	\includegraphics[scale=0.45]{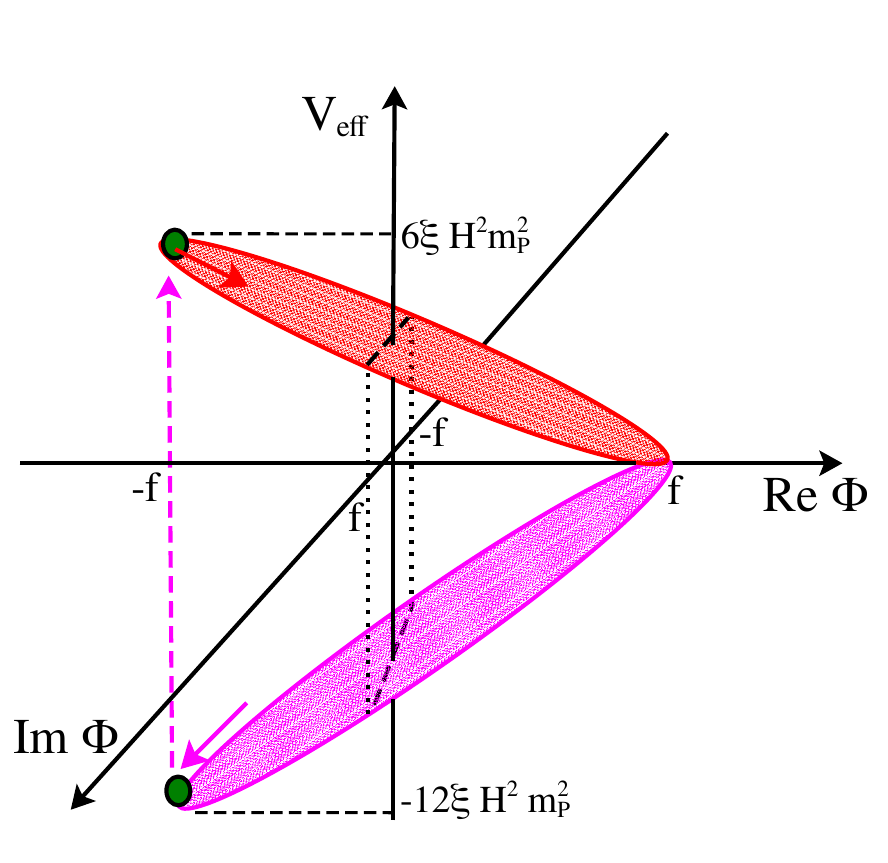}
	\includegraphics[scale=0.45]{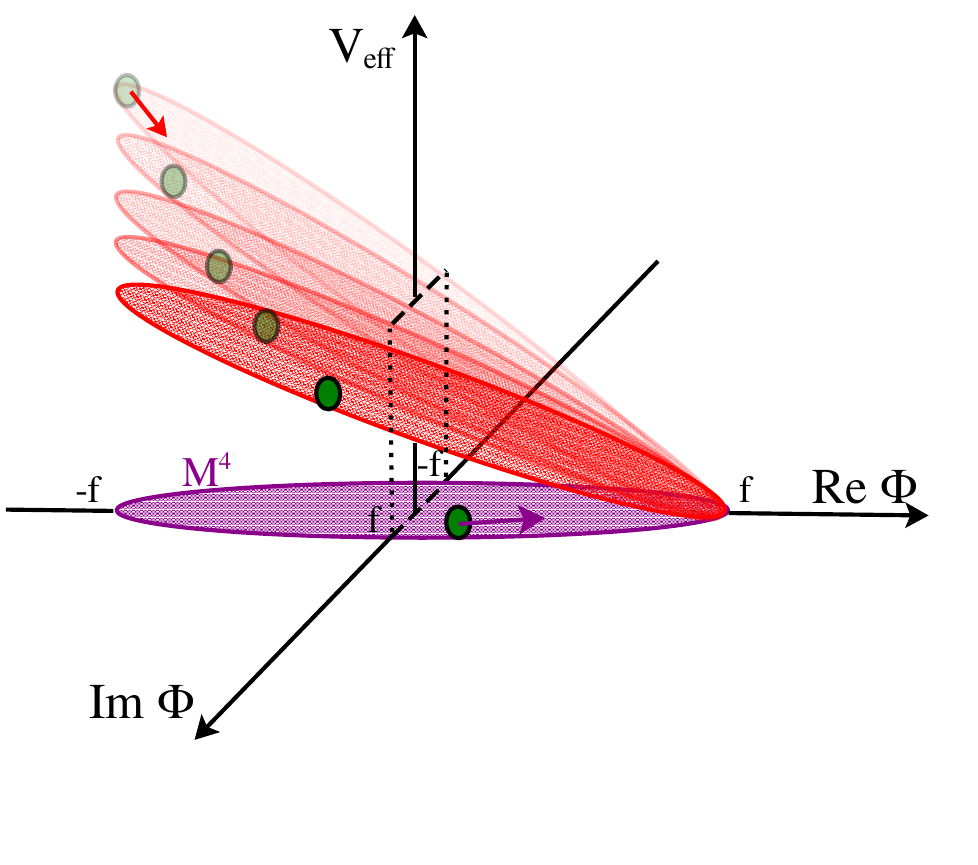}
	\caption{Schematic diagram to visualise 
		the evolution of the axion vacuum manifold.
		{\it Left panel}: Flipping of the tilt of the vacuum manifold at the end of inflation. During inflation, the vacuum manifold is depicted by the purple ellipse. The axion is driven to its minimum at \mbox{$\phi=\pi\,f$}. After inflation, kination begins and the vacuum manifold is tilted in the opposite direction, and it is now depicted by the red ellipse. When the tilt changes, the axion finds itself at the maximum and starts rolling towards the new minimum, which is now at \mbox{$\phi=0$}. {\it Right panel}: Evolution after inflation. While the axion rolls towards the minimum at \mbox{$\phi=0$}, the tilt of the vacuum manifold diminishes with time (red ellipses), in tandem with the axion kinetic energy density. The rotating axion overcomes the potential hill when climbing back the vacuum manifold, because the hill has diminished accordingly. The figure also depicts the disappearing of the tilt after reheating, when $R=0$. The axion continues to rotate around an almost horizontal vacuum manifold (purple ellipse), given that its potential is given by Eq.~\eqref{V}, with $M$ very small.}
\label{fig:tilt}
\end{figure*}

Thus, the tilt of the vacuum manifold is flipped as inflation ends and kination begins, as shown in Fig.~\ref{fig:tilt}.
The axion finds itself at the maximum of the potential, so it rolls down towards the minimum.

We consider that the rolling axion remains subdominant. After inflation, Eq.~\eqref{Veffkin} suggests that the maximum potential density is \mbox{$V_{\rm eff}^{\rm max}=6\xi m_P^2H^2$}. Thus, we require
\begin{equation}
	1\gg\frac{V_{\rm eff}^{\rm max}}{\rho}=2\xi ~,
	\label{ximax}
\end{equation}
where we considered \mbox{$\rho=3m_P^2H^2$}. 

During inflation the axion is massive because
\mbox{$||V_{\rm eff}''||=3\xi m_P^2H^2/f^2>(\frac23 H)^2$}. 
This means that it is driven towards the minimum expectation value \mbox{$\phi=\pi\,f$}. Thus, the range of the non-minimal coupling is
\begin{equation}
	\frac{3}{4}\left(\frac{f}{m_P}\right)^2<\xi\ll \frac{1}{2} ~.
	\label{xirange}
\end{equation}

\begin{figure*}
\begin{center}
	\includegraphics[width=0.5
	\textwidth]{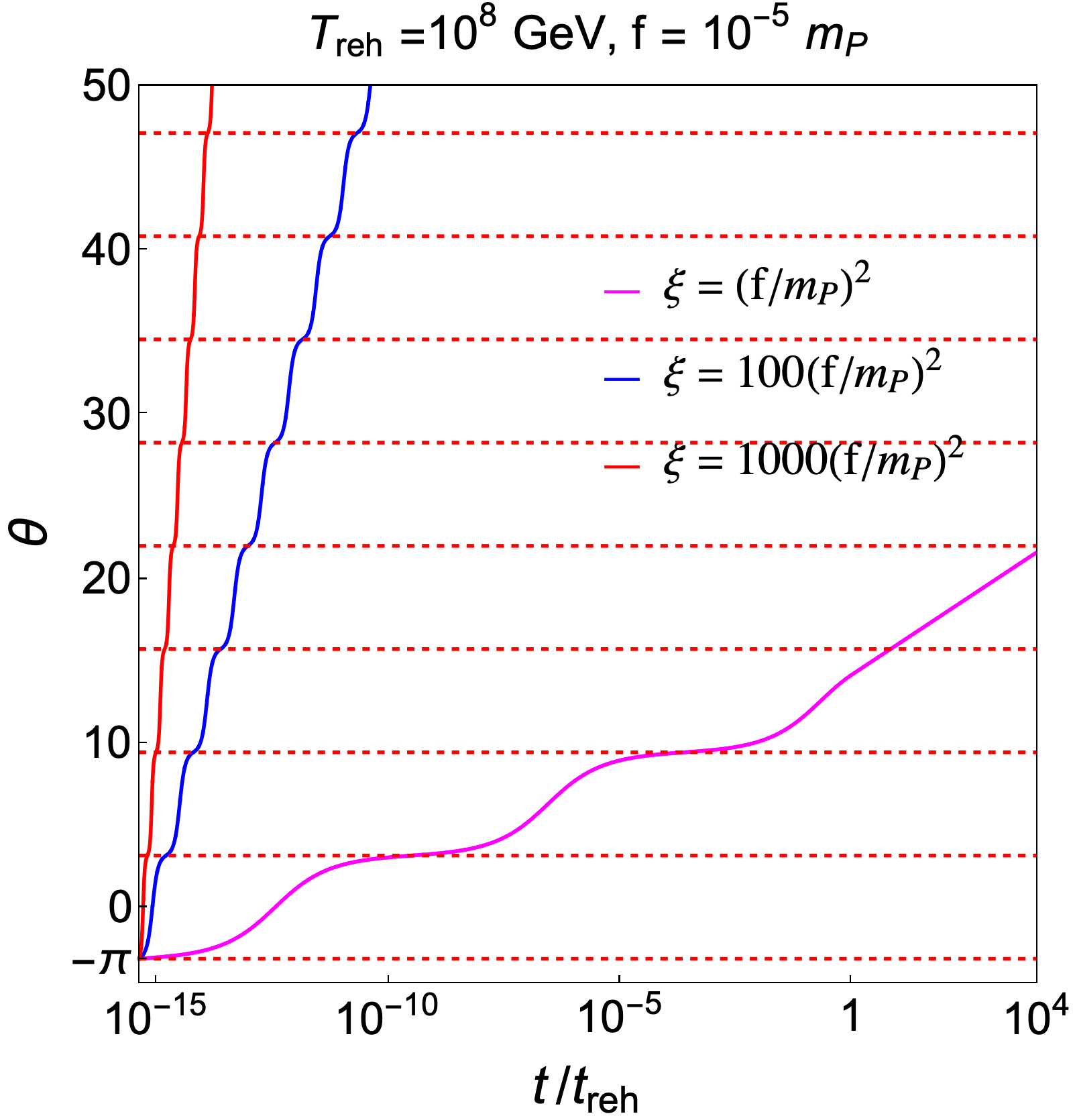}
\end{center}	
	\caption{The rotation of the axion ($\theta$), with the red dashed lines indicating the potential barriers at $-\pi,\pi, 3 \pi, 5 \pi, ...~$. The different colors denote different values of $\xi$, with rotation taking place even for $\xi=\left(f/m_P\right)^2$.}
\label{fig:numthet}
\end{figure*}

After inflation, the axion field manages always to overcome the decreasing potential hill and keeps rotating around the vacuum manifold (see Fig.~\ref{fig:numthet}). 
The equation of motion for $\theta=\phi /f$ is given by 
\begin{align}
	\ddot{\theta}+3H\dot{\theta}+ V'(\phi)/f=0 \,, \label{eq:EOMthet}
\end{align}
where $V'(\phi)/f=\frac{3\xi m_P^2 H^2}{f^2} \sin\theta$. 

\begin{figure*}
	\centering   \includegraphics[scale=0.5]{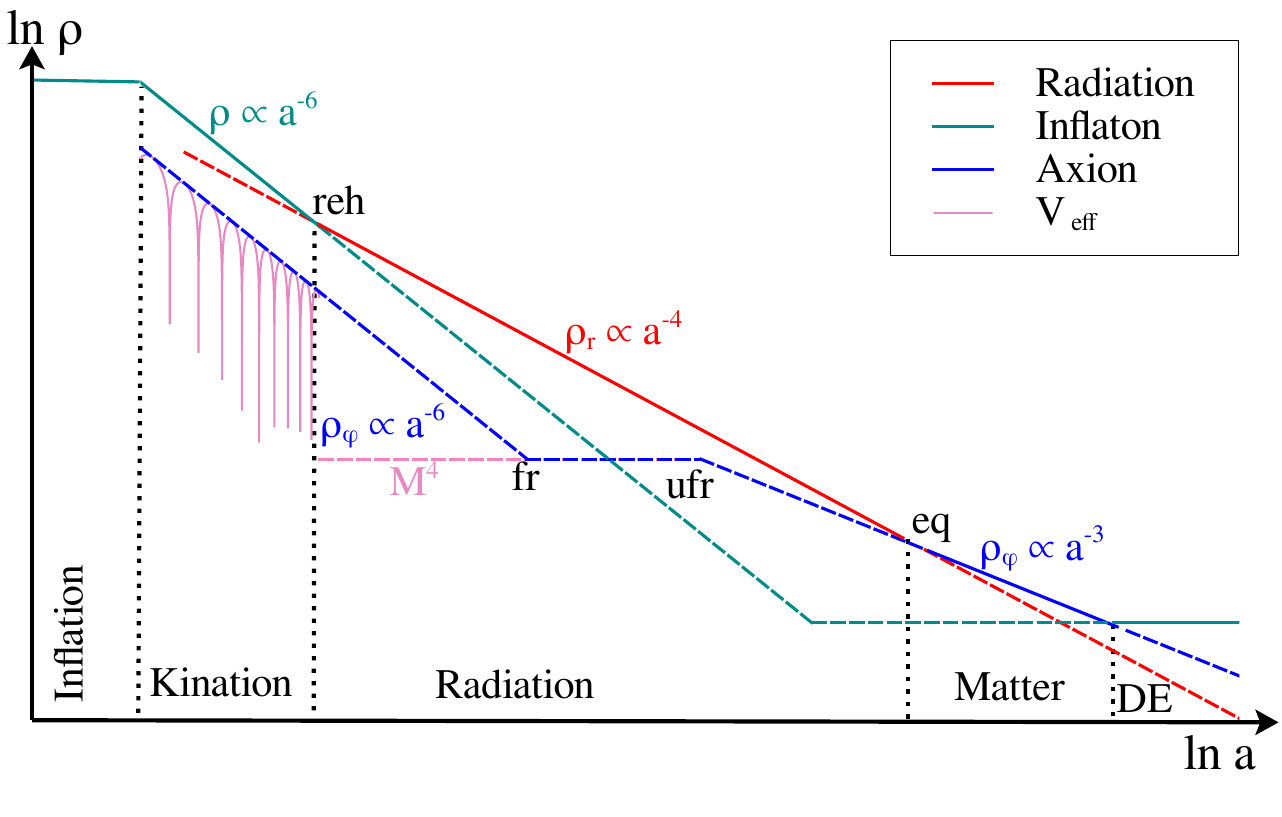}
	\caption{
Log-log plot of the evolution of the energy densities of the different components of the Universe. The solid lines indicate the dominant component. We have assumed a model of quintessential inflation, such that the inflaton condensate gives rise eventually to the dark energy (DE) at present. Note that this is not necessary.  Our mechanism would operate with any non-oscillatory model of inflation followed by a period of kination.}
	\label{fig:sketch}
\end{figure*}

After the end of inflation there is a period of kination, when the density of the Universe scales as \mbox{$\rho\propto a^{-6}$}, while radiation, which appears at the end of inflation, scales as \mbox{$\rho_r\propto a^{-4}$}. 
The rotating axion density remains at constant ratio $\rho_\phi/\rho$ during kination.
After reheating, \mbox{$R=0$} and $V_{\rm eff}$ disappears.  
The axion continues to rotate around its vacuum manifold, with \mbox{$\rho_\phi\propto a^{-6}$}
The density ratio decreases, since 
\mbox{$\rho_\phi/\rho\propto a^{-2}$}
This continues until \mbox{$M^4\simeq\rho_\phi^{\rm fr}$}   when the axion freezes with \mbox{$\phi_{\rm fr}\sim f$}, cf. Fig.~\ref{fig:sketch}.

\section{Axionic Dark Matter}

The axion remains frozen until its mass becomes comparable to 
the decreasing Hubble scale \mbox{$H(t)=2/t$}.
The axion unfreezes when
\mbox{$H_{\rm ufr}\simeq m_\phi\simeq M^2/f$}.
Then, the axion oscillates around its minimum with  
\mbox{$\rho_\phi\propto a^{-3}$} such that \mbox{$\rho_\phi/\rho\propto a$}. The axion dominates at equality
\begin{equation}
	1 \simeq\left.\frac{\rho_\phi}{\rho}\right|_{\rm eq}
	\Rightarrow\;M\sim 
	\sqrt{\frac{m_P}{t_{\rm eq}}}\left(\frac{m_P}{f}\right)^{3/2},
\end{equation}
which suggests
\begin{equation}
	M\sim 10^{-9}\,{\rm GeV}\left(\frac{m_P}{f}\right)^{3/2}.
	\label{M}
\end{equation}

After equality 
\mbox{$R=3H^2$}. However, the oscillating axion around
\mbox{$\phi=0$}
has \mbox{$|\phi|\ll f$}. As a result, gravity remains Einsteinian because 
\mbox{$\gamma^2(\phi)\simeq 1+\frac12\xi(\phi/f)^2\approx 1$}.

\section{Baryogenesis}

The rotating axion can generate the baryon asymmetry of the Universe
\cite{Cohen:1987vi,Cohen:1988kt,Domcke:2020kcp}.  
The required rotation is due to the flipping of the potential at the end of inflation, rather than explicit U(1) breaking operators.
The non-zero velocity $\dot\theta$ spontaneously breaks CPT in the expanding Universe.In the presence of baryon number violating interactions, this  leads to the generation of baryon asymmetry, through spontaneous baryogenesis.

The axion generates a chemical potential \mbox{$\mu\sim\dot\phi/f$} for quarks and/or leptons.
This occurs through derivative couplings of the axion with fermion currents of the form $x_\psi \partial_\mu \phi j^{\mu}_{\psi}/ f$, where $\psi$ indicates a fermion with $B-L$ charge $x_\psi$ and \mbox{$j^{\mu}_{\psi}=\bar{\psi}\gamma_\mu \psi$}.
This coupling accompanied by a B (or L) violating interaction causes an energy shift in particles, giving rise to baryon or lepton asymmetry
\mbox{$n_B\sim n_L\sim\dot{\theta}T^2$}. 
The velocity in field space $\dot\theta$ must be large enough for the observed asymmetry $Y_B$. 

The production of asymmetry continues as long as the baryon or lepton number violating interaction remains in thermal equilibrium.  When such an interaction decouples, say at a temperature $T_{\rm B-L}$, the asymmetry gets frozen to a constant value. In all cases we find
\begin{align}
	Y_B \simeq  \frac{3 \sqrt{30\xi} c_B }{2 \pi \sqrt{g_*} } \frac{ T_{\rm B-L}^2}{ f ~T_{\rm reh}} \,,\label{eq:YB2}
\end{align}
where the reheating temperature is $T_{\rm reh}$, $g_{*}$ is the effective relativistic degrees of freedom and $c_B$ is an $\mathcal{O}(1)$ factor to be calculated from the transport equations. The above suggests
\begin{align}
	4.8 \times 10^{-10}  \left(\frac{f}{\text{GeV}}\right) 
	\ll  \frac{c_B}{\rm GeV} \frac{T_{\rm B-L}^2}{T_{\rm reh}} \lesssim 9.5 \times 10^8\,.
	\label{eq:xirange2}
\end{align}

\section{Primordial Gravitational Waves}

\begin{figure*}[ht]
	\centering
	\includegraphics[width=0.8\textwidth]{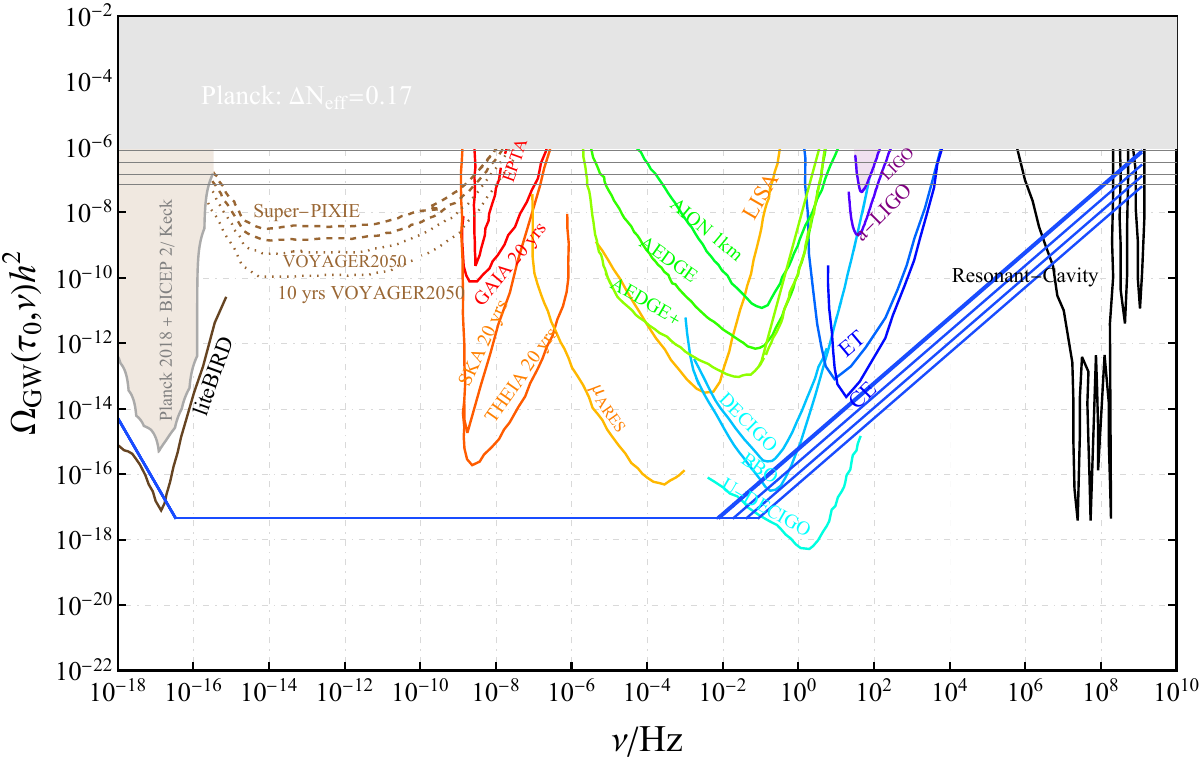}
	\caption{
		The current GW spectrum (blue), choosing various values of reheating temperature $T_{\rm reh}$. The different gray solid lines indicate the future sensitivity reaches of several experiments for $\Delta N_{\rm eff}$, i.e.  BBN+CMB, CMB-S4/PICO, CMB-HD, COrE/EUCLID, and hence to the peak of the GW energy spectra. In the figure, the lowest possible values of $T_{\rm reh}$ are shown. If $T_{\rm reh}$ were even lower (and kination lasted longer) then the peak in the GW spectrum would violate the BBN bound, depicted by the horizontal gray band on top of the~figure.}
	\label{fig:gw}
\end{figure*}

The observed energy spectrum of primordial gravitational waves (GW) generated during inflation is affected by the thermal history after inflation \cite{Gouttenoire:2021jhk,Figueroa:2019paj,Chen:2024roo}. The spectral GW density parameter is 
\begin{equation}
	\Omega_{\rm GW}(\nu) \propto \nu^\beta ~,
	\quad {\rm where} \quad
	\beta = 2\, {w - 1/3 \over w + 1/3} ~,
	\label{fbeta}
\end{equation}
where $\nu$ is the GW frequency.
Then, the observed GW energy spectrum is 
	\begin{equation} \label{eq:omegaGW_1}
		\Omega_{\rm GW}(\tau_0, \nu) \simeq \Omega_{\rm GW}^{\rm RD}
		\left\{
		\begin{matrix}
			&\nu/\nu_{\rm reh}
			~, \quad &&\nu_{\rm reh} < \nu < \nu_{\rm end}
			\\
			&1 ~,  &&\nu_{\rm eq} < \nu < \nu_{\rm reh}
			\\
			&(\nu_{\rm eq}/\nu)^2 ~,  &&\nu_{0} < \nu < \nu_{\rm eq}
		\end{matrix}
		\right. ~,
	\end{equation} 
where $\Omega_{\rm GW}^{\rm RD}$ is a constant representing the GW density parameter of modes that reenter the horizon during RD. For GUT-scale inflation, we have \mbox{$\Omega_{\rm GW}^{\rm RD} \sim 10^{-17}$} (see Fig.~\ref{fig:gw}). 

The GW background acts as an extra radiation component. These GWs can contribute to extra relativistic degrees of freedom during BBN, parametrised by the quantity $\Delta N_{\rm eff}$. The value of $\Delta N_{\rm eff}$ is being measured observationally. Consequently, it can be used to put a constraint on the GW amplitude and subsequently on the value of the reheating temperature \cite{Maggiore:1999vm,Boyle:2007zx,Caprini:2018mtu}: 
\begin{equation}  \label{eq:bbn_int}
	\int_{\nu_{\rm BBN}}^{\nu_{\rm end}}\frac{{\rm d} \nu}{\nu}\Omega_{_{\rm GW}}(\nu) \,h^2 \leq \frac{7}{8}\left(\frac{4}{11}\right)^{4/3}\Omega_{\rm \gamma}h^2\,\Delta N_{\rm eff} ~,
\end{equation}
where $\Omega_{\rm \gamma}h^2\simeq 2.47\times10^{-5}$ corresponds to  the relic density of the radiation measured today, and $\nu_{\rm BBN} \sim 10^{-11} {\rm Hz}$. Using the observed value, \mbox{$\Delta N_{\rm eff}=0.17$}, we find \mbox{$\Omega_{\rm peak} h^2 \lesssim 7.3 \times 10^{-7}$}, which corresponds to the bound 
\begin{equation}
	T_{\rm reh} \gtrsim 2.2 \times 10^{7} ~\text{GeV} \,.
\end{equation}

We can put the above together in an example. Assume GUT-scale inflation \mbox{$H\sim 10^{-5}m_P$} and decay constant \mbox{$f\sim 10^{-2}\,m_P$}. Then, the requirement of successful DM is met when \mbox{$M\sim\,$keV} and \mbox{$m_\phi\sim 10^{-20}\,$eV}. Using \mbox{$T_{\rm reh}\sim 10^7\,$GeV}, the observed baryon asymmetry is generated 
when \mbox{$T_{\rm B-L}\simeq 8\times 10^7\,$GeV} and \mbox{$\xi\sim 10^{-4}$}.

\section{Fragmentation}

\begin{figure*}
	\begin{center}
	\includegraphics[width=0.6
	\textwidth]{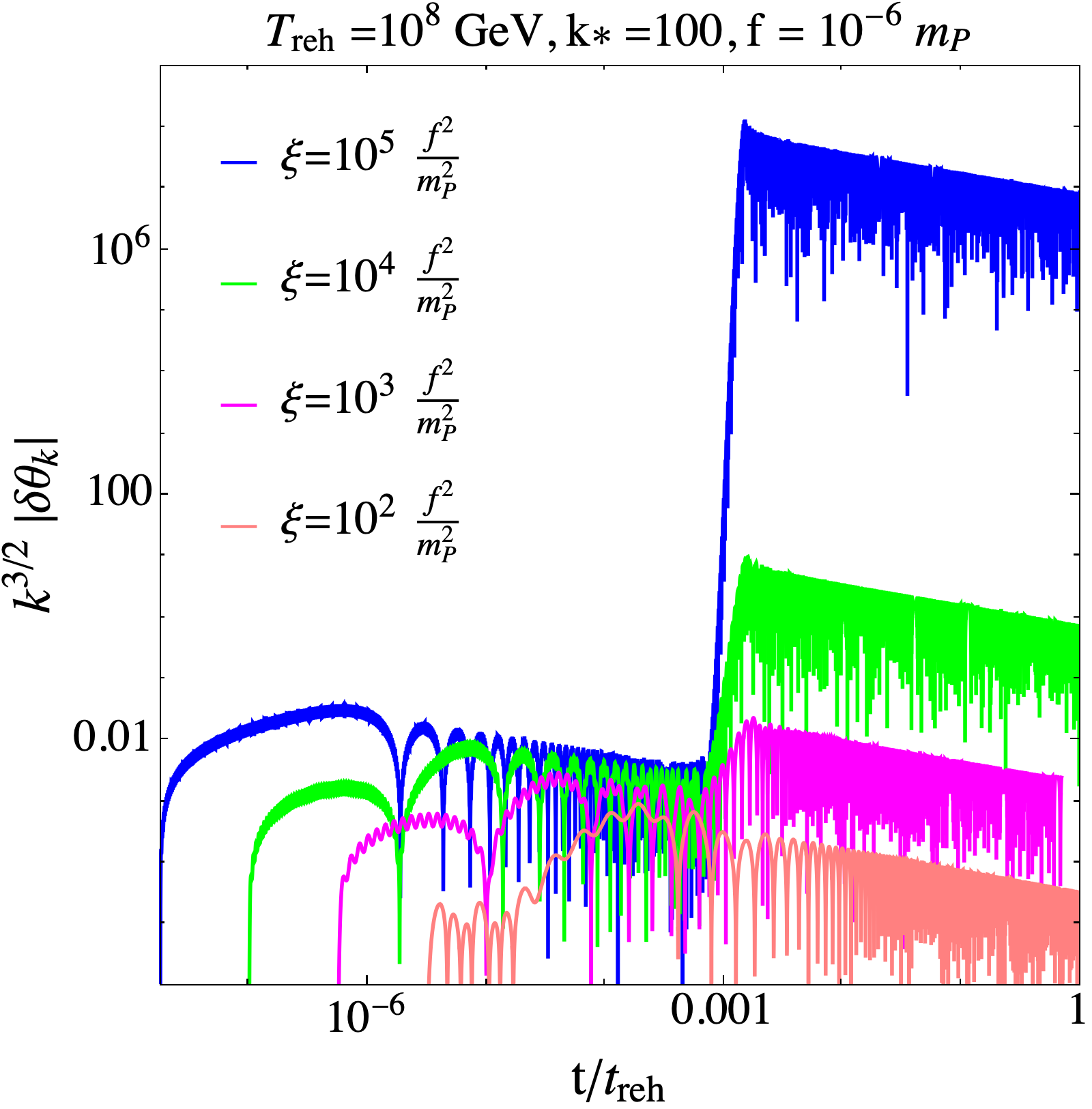}
	\end{center}
	\caption{Evolution of the axion fluctuations for different values of $\xi$, exceeding \mbox{$f^2/m_P^2$}. Substantial growth can be seen with the increase in the value of $\xi$. However, we can see that \mbox{$\xi\sim 10^2(f/m_P)^2$} does not result in resonant amplification.}
	\label{fig:parres}
\end{figure*}

The enhancement of axion fluctuations can lead to a significant loss of kinetic energy of the rotating axion.
This occurs  through parametric resonance and is called axion fragmentation \cite{Fonseca:2019ypl,Eroncel:2022vjg,Eroncel:2022efc}. To study this, the axion is decomposed as
\mbox{$\Theta(t,\textbf{x})= \theta(t) + \delta \theta (t,\textbf{x})$}.
The equation of motion for the Fourier modes $\delta\theta_k(t)$ is
\begin{equation}
	\delta\ddot{\theta}_k+3H\delta \dot{\theta}_k+\left[\frac{k^2}{a^2}+ V''(\phi)\right]\delta\theta_k=0\,,\label{eq:thetkeom}
\end{equation}
where  $V''(\phi)=\frac{3\xi m_P^2 H^2}{f^2} \cos(\frac{\phi}{f})$ with $H=1/3t$ during kination. 
The instability bands are
\begin{equation}
	\frac{\dot{\theta}^2}{4} - \frac{3\xi m_P^2 H^2}{2f^2}\lesssim \frac{k^2}{a^2}\lesssim \frac{\dot{\theta}^2}{4} + \frac{3\xi m_P^2 H^2}{2f^2}\,.\label{eq:insblbnd}
\end{equation}
The maximum growth is around $\exp\left(
	\frac{3\xi m_P^2 H^2}{2f^2\dot{\theta}}t
	\right)$.
Thus, for the modes to grow, we need
\mbox{$H<\frac{3\xi m_P^2 H^2}{2f^2\dot{\theta}}$}.
This gives us the conservative bound 
\begin{equation}
	\xi \lesssim 5.3\left(\frac{f}{m_P}\right)^2\,
	\label{eq:xifragbnd}
\end{equation}
Numerical investigation suggests that this bound is relaxed by at leas an order of magnitude (see Fig.~\ref{fig:parres}).

\section{The Kibble issue}

We consider  an ever-existing axion, during and after inflation, which means that there is no phase transition which gives rise to topological defects
\cite{Kibble:1980mv}.
However, the rotating axion still suffers from the Kibble mechanism problem: after inflation, the axion rotation can be clockwise or anticlockwise, at random.
To avoid the Kibble problem: consider a slight misalignment of the expectation value of the axion at the end of inflation from 
\mbox{$\phi=\pi\,f$}. To this end, assume \mbox{$m_\phi\lesssim H$}. Then \cite{Bunch:1978yq,Mijic:1994vv}
\begin{equation}
	\langle\bar\phi^2\rangle=\frac{3H^4}{8\pi^2 m_\phi^2}\simeq\frac{1}{16\pi^2\xi}\left(\frac{f}{m_P}\right)^2H^2 ~,
	\label{mean}
\end{equation}
where \mbox{$\bar\phi\equiv\phi-\pi\,f$} and we used that near the effective minimum \mbox{$\bar\phi\approx 0$} we have that \mbox{$V_{\rm eff}\simeq\frac12(6\xi m_P^2/f^2)H^2\bar\phi^2$}. To avoid the complete spread of the condensate \mbox{$\langle\bar\phi^2\rangle\ll(\pi\,f)^2$}.
Thus, the range of $\xi$ is now
\begin{equation}
	\frac{1}{16\pi^4}\left(\frac{H}{m_P}\right)^2\ll\xi<\frac14\left(\frac{f}{m_P}\right)^2.
	\label{xirange+}
\end{equation}
Comparing with Eq.~(\ref{xirange}) we find that 
\begin{equation}
	\xi\sim\left(\frac{f}{m_P}\right)^2\,,
	\label{xi}
\end{equation}
i.e. near the edge of both ranges in Eqs.~\eqref{xirange}
and \eqref{xirange+}. For this, we require that $\xi$ is mildly evolving \cite{Hamada:2014wna,Ezquiaga:2017fvi,Drees:2019xpp}
\begin{equation}
	\xi(\sigma)=\xi_0\left[1+\beta\ln\left(\frac{\sigma^2}{\mu^2}+1\right)\right]\,,\label{xiofsigma}
\end{equation}
where $\sigma$ is the inflaton field, $\mu$ is a constant energy scale and $\xi_0,\beta$ are dimensionless constants which depend upon the microscopic details of the model. During slow-roll inflation, since the inflaton is light, $\sigma$ is almost invariant so that $\xi$ is approximately constant, as assumed. During kination, however, $\sigma$
fast-rolls such that \mbox{$\sigma\propto\ln(t/t_{\rm end})$}, so that \mbox{$\xi\propto\ln[\ln(a)]$}, were \mbox{$a\propto t^{1/3}$}. This means that $\xi$ is not changing fast, and considering it approximately constant is a good approximation. Then, in view of 
Eq.~(\ref{eq:YB2}) we find
\begin{equation}
	\frac{T_{\rm B-L}^2}{T_{\rm reh}}\sim Y_B\, m_P\sim 10^8\,{\rm GeV}\,,\label{eq:barpredtn}
\end{equation}
where we used \mbox{$Y_B\sim 10^{-10}$}. 

\section{Conclusions}

We investigated an attractive new way to generate rotation for an ALP particle, if it is non-minimally coupled to gravity, respecting the shift-symmetry, without introducing an explicit U(1) breaking operator. The rotation is generated when the effective potential of the axion flips at the end of inflation, which is possible if inflation is followed by kination.

This rotation can generate the baryon asymmetry $Y_B$ through spontaneous baryogenesis determined crucially by the non-minimal coupling $\xi$ and the decay constant $f$, and the reheating temperature $T_{\rm reh}$. $Y_B$ is also determined by the decoupling temperature of the baryon or lepton number violating interaction $T_{\rm B-L}$, which depends on the details of the particle physics setup. The axion continues to rotate after reheating until the kinetic energy becomes comparable to the height of the bare mass potential $M^4$. Afterwards the coherent oscillation of the axion behaves as dark matter. We also investigated the danger of the fragmentation of the axion condensate as well as effects of the Kibble issue.

We find that the condition for success is \mbox{$\xi\sim(f/m_P)^2$}. Then, successful baryogenesis requires \mbox{$\frac{T_{\rm B-L}^2}{T_{\rm reh}}\sim Y_B\, m_P\sim 10^8\,{\rm GeV}$} and successful dark matter needs \mbox{$M\sim 10^{-9}\,{\rm GeV}\left(\frac{m_P}{f}\right)^{3/2}$}, while the axion mass is \mbox{$m_\phi\sim M^2/f$}. For example, for GUT-scale inflation \mbox{$H\sim 10^{-5}\,m_P$} and decay constant \mbox{$f\sim 10^{-2}\,m_P$} we have \mbox{$\xi\sim 10^{-4}$}, which results in \mbox{$M\sim\,$keV} and \mbox{$m_\phi\sim 10^{-20}\,$eV}. Considering also \mbox{$T_{\rm reh}\sim 10^7\,$GeV} we obtain \mbox{$T_{\rm B-L}\simeq 8\times 10^7\,$GeV}.

The scenario is very predictive and favours specific values of the non-minimal coupling $\xi$. These predictions can be indirectly tested by GW experiments like BBO, U-DECIGO and resonant cavities.

\section*{Acknowledgements}

KD is supported (in part) by the STFC consolidated Grant: ST/X000621/1.

\end{document}